\input{aipcheck}

\documentclass[
    ,final            
  ]
  {aipproc}

\layoutstyle{8x11single}

\begin{document}

\title{Detector Development for a Sterile Neutrino Search with the KATRIN Experiment}



\author{Tim Brunst}{address={Max Planck Institute for Physics, F\"ohringer Ring 6, 80805 M\"unchen, Germany}}

\author{Konrad Altenm\"uller}{address={Technical University of Munich, Arcisstr. 21, 80333 M\"unchen, Germany}}

\author{Tobias Bode}{address={Max Planck Institute for Physics, F\"ohringer Ring 6, 80805 M\"unchen, Germany}}

\author{Luca Bombelli}{address={XGLab srl, Bruker Nano Analytics, Via Conte Rosso 23, 20134 Milano, Italy}}

\author{Vasiliy Chernov}{address={Institute for Nuclear Research of Russian Academy of Sciences, Prospekt 60-letiya Oktyabrya 7a, Moscow 117312, Russian Federation}}

\author{Anton Huber}{address={Karlsruhe Institute of Technology, Hermann-von-Helmholtz-Platz 1, 76344 Eggenstein-Leopoldshafen, Germany}}

\author{Marc Korzeczek}{address={Karlsruhe Institute of Technology, Hermann-von-Helmholtz-Platz 1, 76344 Eggenstein-Leopoldshafen, Germany}}

\author{Thierry Lasserre}{address={Centre CEA de Saclay, 91191 Gif-sur-Yvette cedex, France}, altaddress={Technical University of Munich, Arcisstr. 21, 80333 M\"unchen, Germany}}

\author{Peter Lechner}{address={Halbleiterlabor of the Max Planck Society, Otto-Hahn-Ring 6, 81739 M\"unchen, Germany}}

\author{Susanne Mertens}{address={Max Planck Institute for Physics, F\"ohringer Ring 6, 80805 M\"unchen, Germany}, altaddress={Technical University of Munich, Arcisstr. 21, 80333 M\"unchen, Germany}}

\author{Aleksander Nozik}{address={Institute for Nuclear Research of Russian Academy of Sciences, Prospekt 60-letiya Oktyabrya 7a, Moscow 117312, Russian Federation}, altaddress={Moscow Institute of Physics and Technology, 9 Institutskiy per., Dolgoprudny, Moscow Region, 141700, Russian Federation}}

\author{Vladislav Pantuev}{address={Institute for Nuclear Research of Russian Academy of Sciences, Prospekt 60-letiya Oktyabrya 7a, Moscow 117312, Russian Federation}}

\author{Daniel Siegmann}{address={Max Planck Institute for Physics, F\"ohringer Ring 6, 80805 M\"unchen, Germany}}

\author{Aino Skasyrskaya}{address={Institute for Nuclear Research of Russian Academy of Sciences, Prospekt 60-letiya Oktyabrya 7a, Moscow 117312, Russian Federation}}

\begin{abstract}
The KATRIN (Karlsruhe Tritium Neutrino) experiment investigates the energetic endpoint of the tritium $\beta$-decay spectrum to determine the effective mass of the electron anti-neutrino with a precision of $200\,\mathrm{meV}$ ($90\,\%$ C.L.) after an effective data taking time of three years.

The TRISTAN (tritium $\beta$-decay to search for sterile neutrinos) group aims to detect a sterile neutrino signature by measuring the entire tritium $\beta$-decay spectrum with an upgraded KATRIN system. One of the greatest challenges is to handle the high signal rates generated by the strong activity of the KATRIN tritium source. Therefore, a novel multi-pixel silicon drift detector is being designed, which is able to handle rates up to $10^{8}\,\mathrm{cps}$ with an excellent energy resolution of $<200\,\mathrm{eV}$ (FWHM) at $10\,\mathrm{keV}$.

This work gives an overview of the ongoing detector development and test results of the first seven pixel prototype detectors.
\end{abstract}

\maketitle


\section{Introduction}

\paragraph{Sterile neutrinos}
The nature of dark matter in our universe is still unknown. Despite overwhelming cosmological evidence for its existence, a direct detection is pending. One promising candidate is a keV-scale sterile neutrino \cite{whitepaper}. Several theories predict sterile neutrinos within a minimal extension of the Standard Model of particle physics (SM). These fermions would not even interact weakly, but would mix with active neutrinos. TRISTAN, as an extension of the KATRIN experiment, could therefore be able to discover such a neutrino by observing the signature of the heavy mass-eigenstate mixing in the tritium $\beta$-decay spectrum.

\begin{figure}
  \includegraphics[height=.2\textheight]{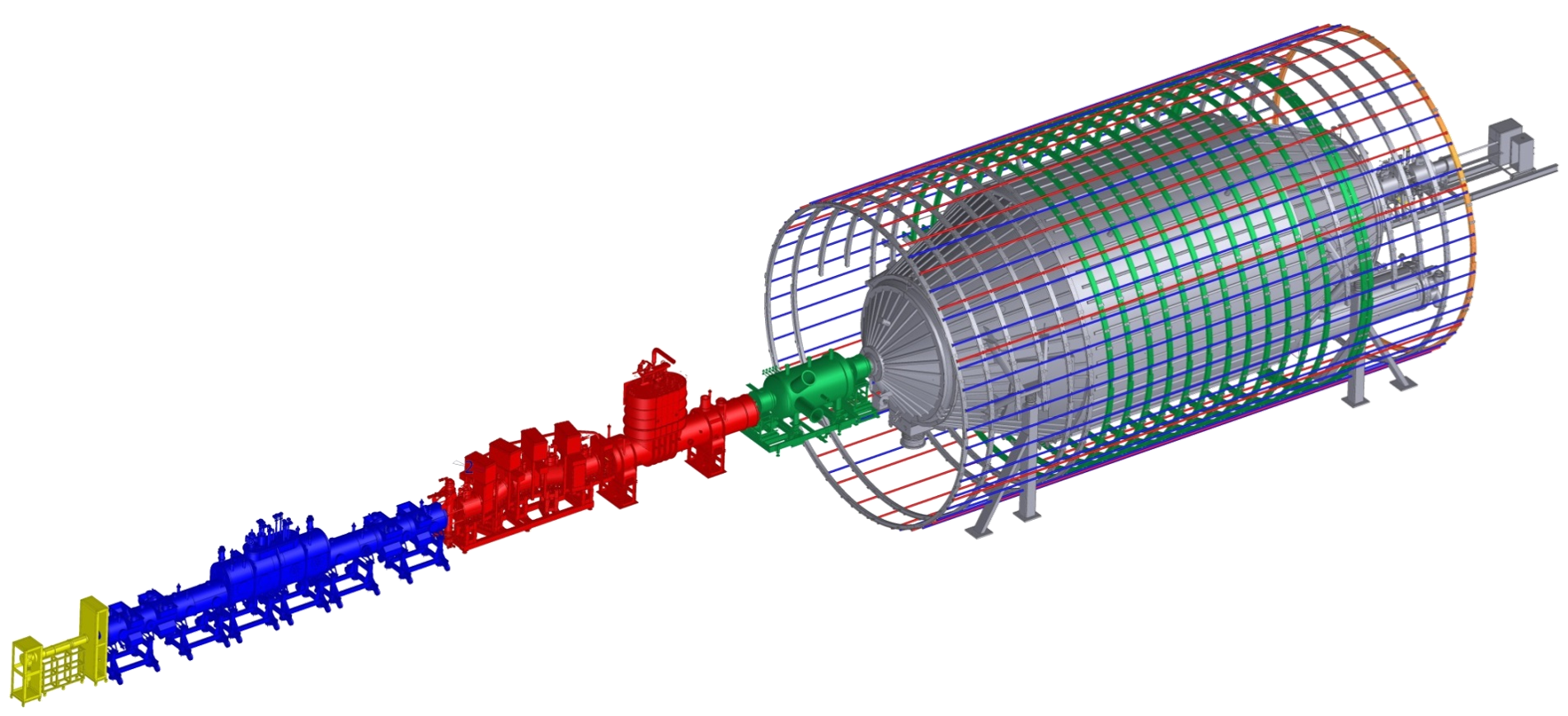}
  \caption{The KATRIN experiment consists of rear wall (\textit{yellow}), WGTS (\textit{blue}), pumping section (\textit{red}), pre-spectrometer (\textit{green}), main spectrometer and detector section (both \textit{grey}).}
  \label{fig:KATRIN}
\end{figure}

\paragraph{The Karlsruhe Tritium Neutrino experiment}
Starting mid 2018 the Karlsruhe Tritium Neutrino (KATRIN) experiment \cite{designreport} will investigate the endpoint region of the tritium $\beta$-decay spectrum in order to determine the absolute neutrino mass (see figure \ref{fig:KATRIN}).
KATRIN combines a high-activity gaseous tritium source with a high-resolution spectrometer. First, tritium gas decays in a windowless gaseous tritium source (WGTS). The electrons are adiabatically guided by magnetic fields throughout the whole setup. The tritium is pumped out by turbo molecular and cryogenic pumps and then recycled. Using the MAC-E filter technique, only $\beta$-electrons that pass a certain energy threshold pass the pre- and main spectrometer and are counted by a detector. The rear wall collects all electrons that do not reach the detector and defines the ground potential.

Unlike the neutrino mass, which manifests itself only at the endpoint region of the $\beta$-decay, the signature of a keV-scale sterile neutrino may appear anywhere in the spectrum depending on the exact mass of the sterile neutrino. Accordingly, the TRISTAN project aims to detect the entire tritium $\beta$-decay spectrum. In this case, electrons of all energies shall pass to the detector. Since the KATRIN detector is not designed to handle such high count rates, a novel detector has to be constructed.

\section{A novel detector}
The novel detector has to fulfill the following requirements:

\begin{description}
\item[Excellent energy resolution:]
In order to be sensitive to the sterile neutrino signal in a differential measurement of the electron energy, an excellent energy resolution of the detector ($300\,\mathrm{eV}$ full width at half maximum (FWHM) @ $20\,\mathrm{keV}$) and a low detection threshold ($1\,\mathrm{keV}$) are required. To this end, a dead layer thickness of less than $100\,\mathrm{nm}$ is necessary.

\item[Handling of high rates:]
In order to reach a statistical sensitivity of $10^{-6}$ in three years a signal rate of $10^8\,\mathrm{cps}$ is necessary. A count rate in the order of $10^5\,\mathrm{cps}$ is manageable for modern data acquisition systems. Thus the detector has to be segmented into $\sim10^3$ pixels with a count rate of $10^5$ each.

\item[Large area coverage:]
To mitigate charge sharing from neighboring pixels the pixel diameter should not be smaller than $2\,\mathrm{mm}$. In order to maintain a high energy resolution even with these large pixels, a small capacitance for each pixel is required.
\end{description}

These requirements point to a silicon drift detector (SDD) array design.

\subsection{The silicon drift detector}
The silicon drift detector measures the amount of ionization of incoming particles in the detector material. The cathode covers the complete entrance window. The point-like anode, however, provides small capacitance for each pixel and thus low noise at short shaping times, i.e. at high rates. Each pixel is equipped with a number of drift rings operated at different potentials which facilitate the transport of the signal charge carriers to the anode. At the same time, these rings divide the monolithic detector chip into pixels reducing the dead area per chip to a negligible level. 

\subsection{First prototypes}
\begin{figure}
  \includegraphics[height=.2\textheight]{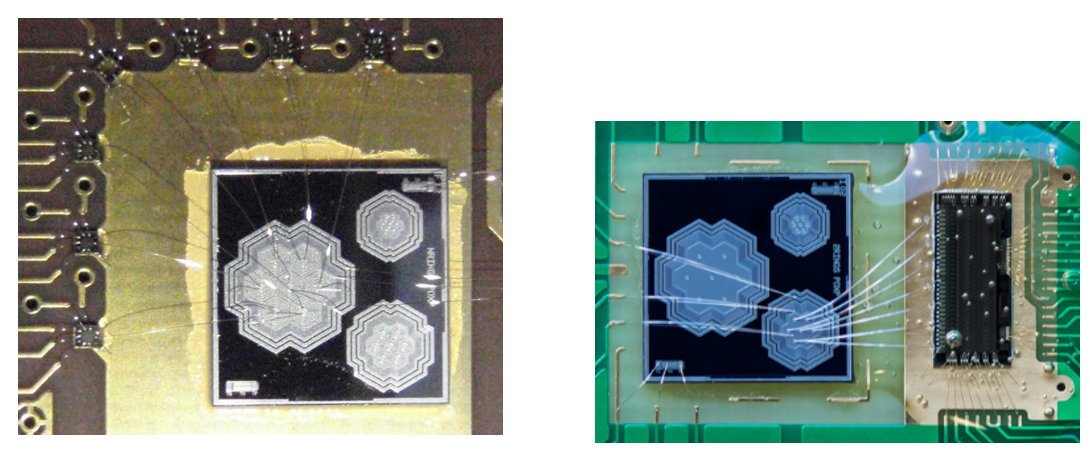}
  \caption{Seven pixel prototype detectors of different sizes equipped with XGLab (\textit{left}) and CEA Saclay (\textit{right}) read-out electronics.}
  \label{fig:detectors}
\end{figure}

Several silicon drift detector prototype chips with seven hexagonal pixels in different sizes ($0.5$, $1$ and $2\,\mathrm{mm}$ in diameter) and a thickness of $450\,\mathrm{\mu m}$ have been produced at the semiconductor laboratory of the Max Planck society in Munich\footnote{https://www.hll.mpg.de/}. They were equipped with two different read-out systems from XGLab, Italy, and CEA Saclay, France (see figure \ref{fig:detectors}).
In both cases, the measured charge is transferred into a voltage by a charge sensitive amplifier (CSA) placed on an application specific integrated circuit (ASIC) for further processing. The XGLab system is optimized for 1-channel SDD detectors, whereas the CEA system allows to operate all channels synchronously. By comparing two different systems, the influence of the read-out electronics on the measured signals and hence systematic effects can be investigated.

\subsection{Characterization}
\begin{figure}
  \includegraphics[height=.2\textheight]{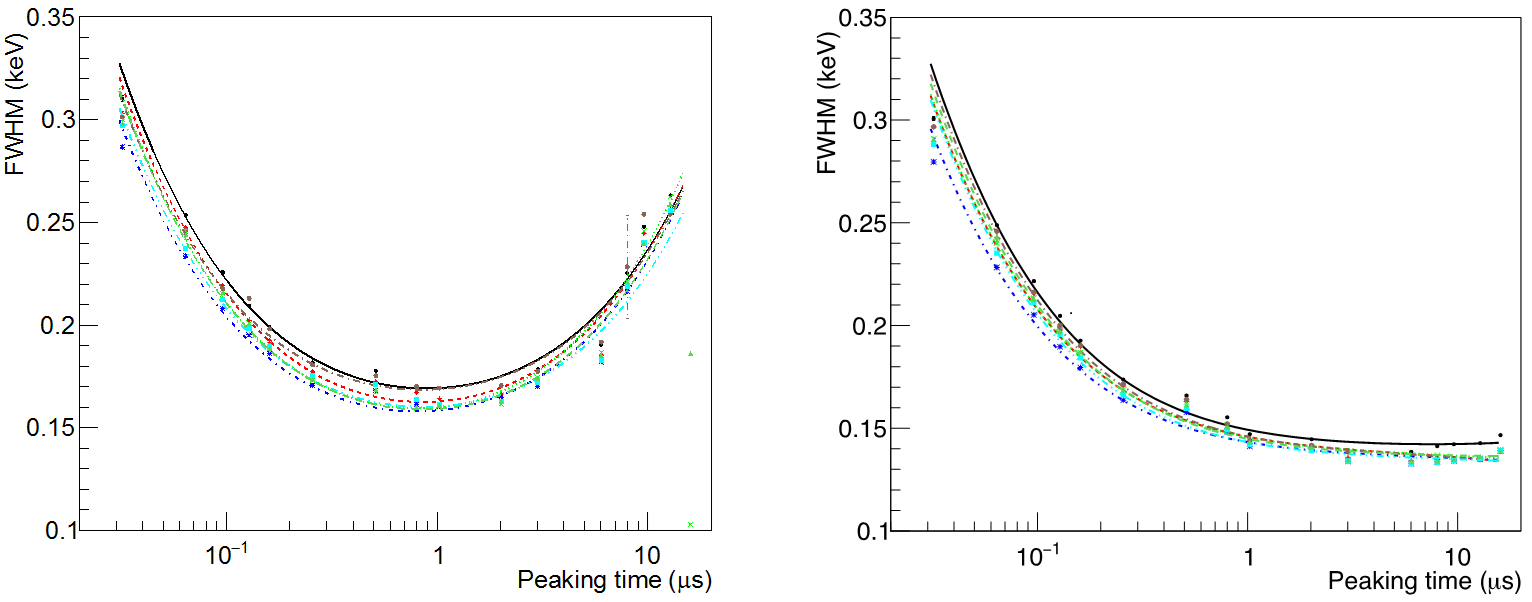}
  \caption{Noise curve at room temperature (\textit{left}) and at $-30\,^\circ\mathrm{C}$ (\textit{right}) for all seven pixels. The noise contribution of leakage current is reduced at low temperature allowing for longer peaking times with better energy resolution.}
  \label{fig:noisecurves}
\end{figure}

The characterization of the systems was performed with radioactive $^{55}\mathrm{Fe}$ and $^{241}\mathrm{Am}$ $\gamma$-sources (among others). The FWHM was measured as a function of peaking time at different temperatures (see figure \ref{fig:noisecurves}). The peaking time is defined as the time from $5\,\%$ of the maximum to the maximum of the shaped signal. The best energy resolution for $2\,\mathrm{mm}$ pixel detectors was obtained with a peaking time of $0.8\,\mathrm{\mu s}$ at room temperature ($163\,\mathrm{eV}$ FWHM) and $2\,\mathrm{\mu s}$ at $-30\,^\circ\mathrm{C}$ ($142\,\mathrm{eV}$ FWHM) respectively.

\subsection{Systematic effects}
\begin{figure}
  \includegraphics[height=.2\textheight]{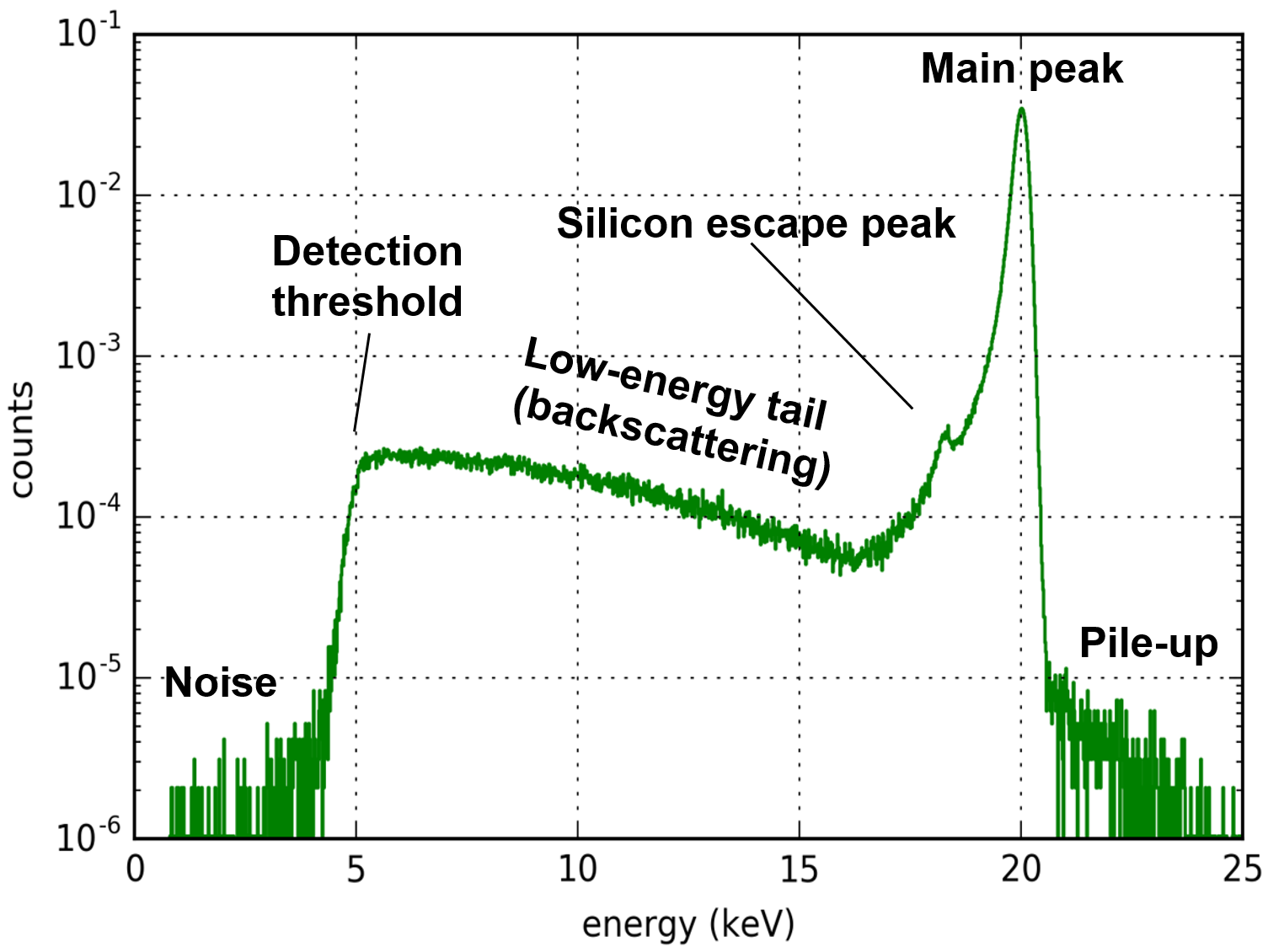}
  \caption{Spectrum of mono-energetic electrons with $20\,\mathrm{keV}$. Clearly visible are effects from pile-up, silicon escape energy, back-scattering, detection threshold and noise. Note the logarithmic y-axis.}
  \label{fig:systeffects}
\end{figure}

The final setup is being designed to measure electrons in the keV-range. These show a different response to the detector than photons. As electrons are massive particles, the systematic effects, such as the influence of the dead layer, back-scattering, charge sharing and others play a major role and are currently investigated. The effect of some of these systematics is depicted in figure \ref{fig:systeffects}.

\begin{description}
\item[entrance window:]
The dead layer is a layer on the detector entrance window which consists of an unavoidable $\mathrm{SiO_2}$ oxidation layer ($\sim10\,\mathrm{nm}$) created by contact with air. Furthermore, the $\mathrm{p}^+$ implantation needed for applying the back-contact voltage prevents the electric field to reach the very surface of the detector. Charge carriers that are generated in this layer are not efficiently drifted to the read-out anode. Unlike photons, electrons are not absorbed in their first interaction but continuously lose energy on their way through the detector material. This leads to a certain amount of energy that is always lost in the dead layer.

\item[back-scattering:]
Electrons have a certain probability to enter the detector, deposit parts of their energy in scattering processes and leave the detector again through the entrance window. Their remaining energy is left undetected. This leads to a low-energy back-scattering background in the measured spectrum.

\item[charge sharing:]
If an electron hits the detector at the border between two adjacent pixels, the created charge cloud is divided by the electric fields and drifted to the respective anodes. This creates a signal in both pixels with the corresponding part of the initial electron energy. If the effect is well understood, the initial energy can be reconstructed as the sum of energies detected in both pixels.

\item[pile-up:]
It can happen that two electrons arrive at one pixel within the signal rise time of around $30\,\mathrm{ns}$. This time difference is too short for the read-out electronics to distinguish between the two events. They will be counted as one event with the sum of the energy of both electrons. For a mono-energetic source this leads to a second line in the spectrum at two times the energy.
\end{description}

\subsection{First tritium data}
\begin{figure}
  \includegraphics[height=.2\textheight]{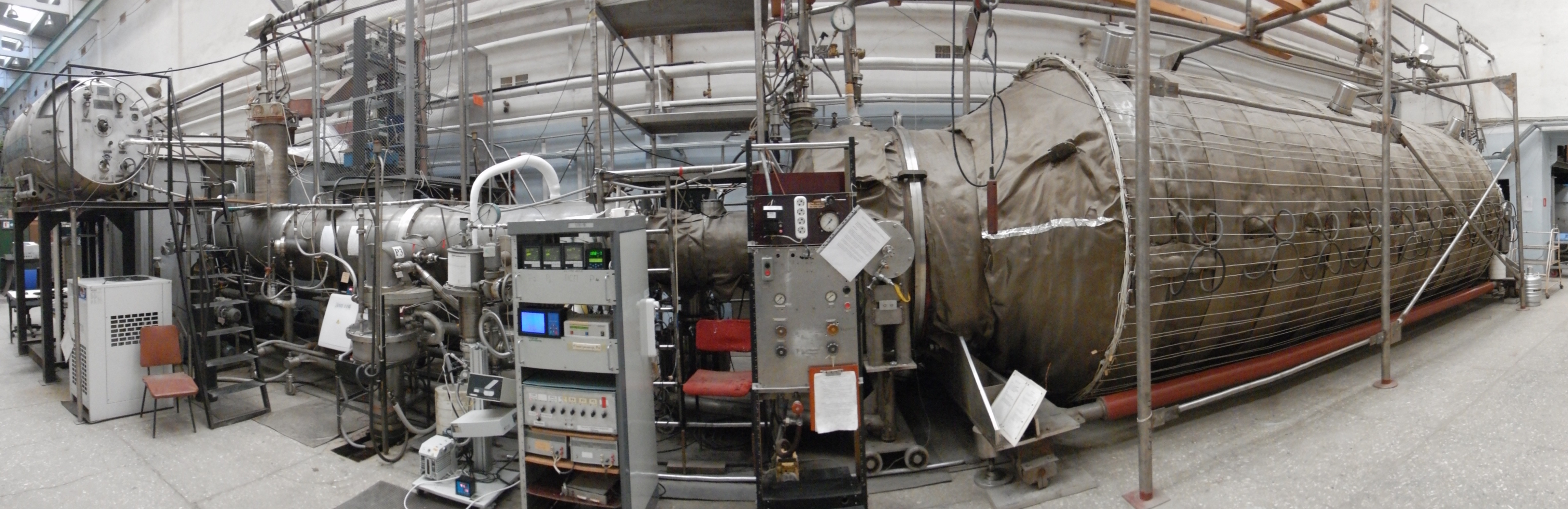}
  \caption{The Troitsk $\nu$-mass experiment holds the world leading limit on the neutrino mass from a direct measurement \cite{troitsklimit}. The collaboration is now searching for a kev-range sterile neutrino.}
  \label{fig:Troitsk}
\end{figure}

\begin{figure}
  \includegraphics[height=.2\textheight]{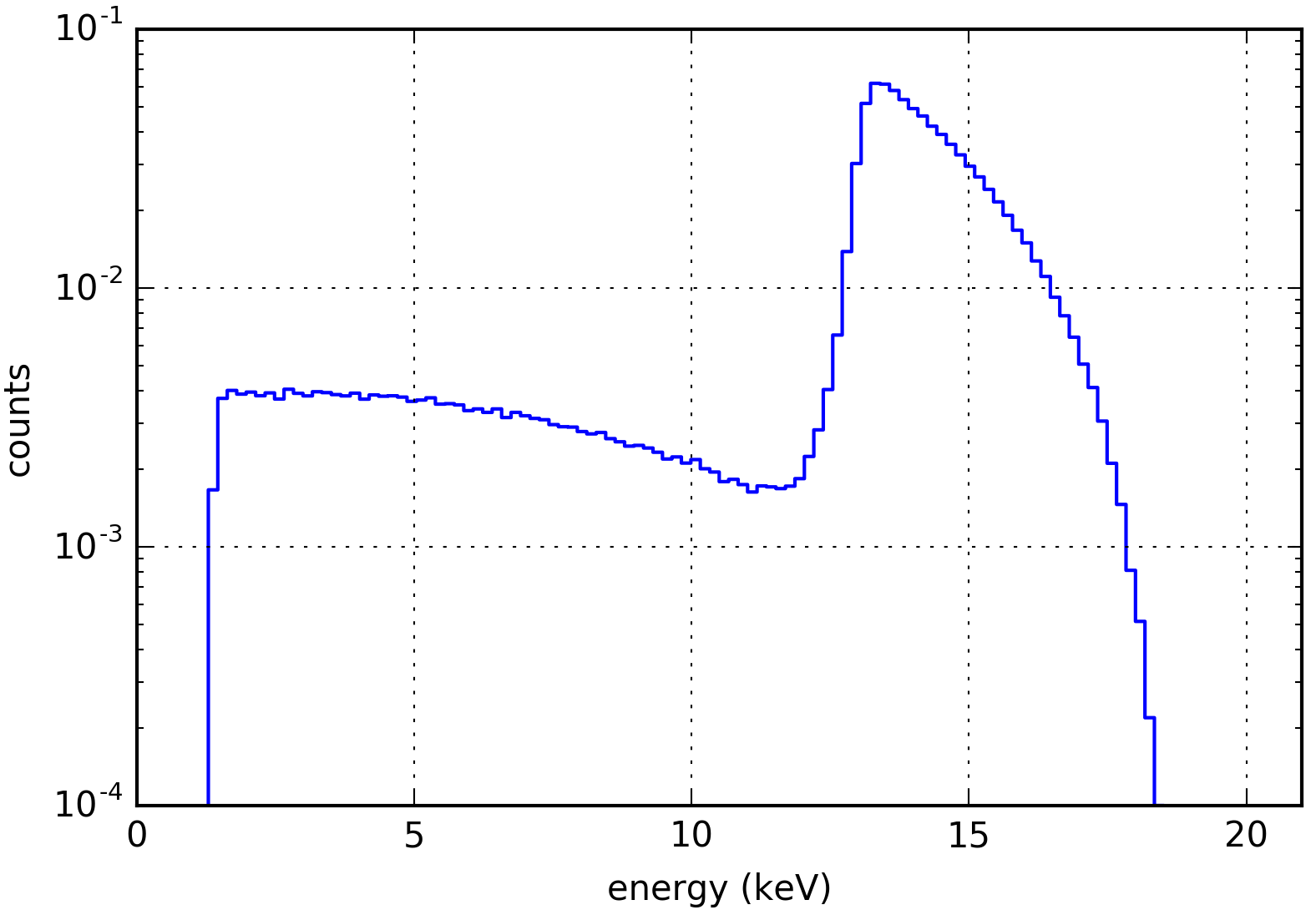}
  \caption{A differential tritium spectrum measured at the Troitsk $\nu$-mass experiment. The retarding potential of the spectrometer was set to $13\,\mathrm{keV}$. The back-scattering background is increased by electrons that were back-reflected at the retarding potential or magnetic mirrors in the setup.}
  \label{fig:tritium}
\end{figure}

In May/June 2017 a $1\,\mathrm{mm}$ prototype detector with CEA read-out electronics was installed at the Troitsk $\nu$-mass experiment (see figure \ref{fig:Troitsk}).
As one of KATRIN's technological predecessor it offers the possibility to investigate systematic effects in a real experimental environment. First tritium data were taken (see figure \ref{fig:tritium}) as well as mono-energetic electron data from the inner electrode system and an electron gun.
This not only allowed for further detector characterization, but also for testing the complete TRISTAN analysis chain including spectral fitting, background control and so forth.

\section{Conclusion and Outlook}
The TRISTAN group aims to detect a sterile neutrino signature by measuring the tritium spectrum with an upgraded KATRIN system. Therefore, a novel silicon drift detector system is being developed. First prototypes have been produced. Their functionality has been successfully demonstrated and first detailed measurements to study systematic effects have been performed.

The characterization of the seven-pixel prototype detectors continues with electron and photon source of different types, e.g. an evaporated Rubidium/Krypton source. A second measurement at the Troitsk $\nu$-mass experiment will be performed with the XGLab read-out electronics system in November 2017. The production of the next prototype generation will start in the beginning of 2018 as an array of 166 pixels with $3\,\mathrm{mm}$ pixel diameter and integrated JFETs as a first signal amplification stage. The final detector is planned to be an array of $~20$ of such detector modules. It will be applied to the KATRIN setup in approximately five years when the neutrino mass measurements are completed.

%

%


\begin{theacknowledgments}
This work was supported by the Max Planck Society (MPG) and the Technical University Munich (TUM). We would like to thank the Institute for Nuclear Research of the Russian Academy of Sciences (INR RAS) and the Halbleiterlabor der Max-Planck-Gesellschaft for the fruitful cooperation. Tim Brunst would like to thank the Joint Institute for Nuclear Research (JINR) and the Czech Technical University (CTU) for their hospitality.
\end{theacknowledgments}



\bibliographystyle{aipproc}   



\end{document}